          [2001/04/25 1.1 (PWD)]
\documentclass[a4paper,twocolumn]{esapub} 
\usepackage{natbib}
\usepackage{epsfig}
\usepackage{graphicx}
\title{SPI observations of positron annihilation radiation from the 4th galactic quadrant: Spectroscopy}
\author{V. Lonjou}
\author{G. Weidenspointner\footnote{ESA Fellow}}
\author{J. Kn{\"o}dlseder}
\author{P. Jean}
\author{M. Allain}
\author{P. von Ballmoos}
\author{M.J. Harris}
\author{J.P. Roques}
\author{G.K. Skinner}
\affil{Centre d'{\'E}tude Spatiale des Rayonnements, BP 4346, 31028
Toulouse, France} 
\author{B.J. Teegarden}
\author{N. Gehrels}
\affil{NASA/GSFC, Code 661, Greenbelt, MD 20771, USA}
\author{N. Guessoum}
\affil{American University of Sharjah, Physics Department, Sharjah, UAE}
\author{C. Chapuis}
\author{Ph. Durouchoux}
\affil{CEA Saclay, 91191 Gif-sur-Yvette, France}
\author{E. Cisana}
\author{M. Valsesia}
\affil{IASF, CNR, 20133 Milan, Italy and Universita degli studi di Pavia, 27100 Pavia, Italy}

\begin{document}


\maketitle

\begin{abstract}The status of the analysis of the electron/positron annihilation radiation performed by INTEGRAL/SPI in the 4th GALACTIC QUADRANT is reported. We use data from the first two Galactic Center Deep Exposures (GCDE) and from the Galactic Plane Scans (GPS). The analysis presented here is focused on the spectroscopic aspects of the electron/positron annihilation radiation. Background substraction and model fitting methods are described, and the parameters of the 511 keV line (flux, energy, and width) are deduced.
\end{abstract}


\section{Introduction}
\label{intro}

The 511 keV line from the Galactic Center (GC) has been observed since the early seventies by a large number of balloon and satellite borne experiments. The first detection was reported in 1972 \citep{johnson72} but at an energy of 426$\pm$26 keV, making the identification of the electron/positron annihilation line uncertain. In 1977, it was observed with a high energy-resolution Germanium spectrometer establishing the line at 511 keV  with a width of only a few keV \citep{leventhal78, albernhe81}.
During the eighties variability in the measured flux was reported after a series of measurements by balloon-borne germanium detectors. The fluctuating results were interpreted as the signature of a compact source of annihilation radiation at the GC \citep{leventhal91}.
Yet during the early nineties this interpretation came into question, since neither eight years of SMM data \citep{share90} nor the revised data of the HEAO-3 Ge detectors \citep{mahoney94} showed evidence for variability in the 511 keV flux. 
Later, during the nineties, CGRO's Oriented Scintillation Spectrometer Experiment (OSSE) measured steady fluxes from the galactic bulge and disk component. The separation was achieved using data from the OSSE, SMM and TGRS experiments \citep{purcell97, milne01}. 
A possible additional component at positive latitude was attributed to an annihilation ``fountain'' in the GC \citep{dermer_skibo97}. Recently, SPI also provided its first results \citep{jean03a, knodlseder03}, giving constraints both on morphology and spectrocopy. A summary of the key parameters measured by high resolution spectrometers is given in Figure~\ref{fig:comparaison}.

We report here the measurement of the Galatic 511~keV gamma ray line emission after one year of INTEGRAL observations, focusing on the spectroscopy aspects. The morphological analysis of the 511 keV emission, using the same data, is presented in \citet{weidenspointner04c}. Both spectroscopy and morphology results were presented during this meeting in the context of previous results (\citet{jean04}). Their consequences on the properties of the annihiliation medium were exposed by \citet{guessoum04}.

\section{Instrument and Data Analysis}
\label{data_analysis}

The Spectrometer for INTEGRAL (SPI) is one of the two main instruments on board the ESA INTEGRAL mission. A detailed description of SPI can be found in \citet{vedrenne03}. SPI is a coded-mask telescope with a fully-coded field of view of $16^\circ$ that uses a 19 pixel high-resolution Ge detector array resulting in an effective area of about 75 cm$^2$ at~511 keV. The energy resolution of the whole camera is 2.10$\pm$0.02 keV at 511 keV and varies between 2 and 8 keV over the full energy range (20 -- 8000 keV). 

The data analyzed in this work were accumulated during the first year's GCDE and GPS, executed as part of INTEGRAL's guaranteed time observations. Both the GCDE and the GPS observations are part of the so-called core program observations of the INTEGRAL observatory, which are proprietary to the INTEGRAL Science Working Team (ISWT) and the instrument teams for one year \citep{winkler01}. A total exposure time of 3475 ks has been obtained during two different periods March-April 2003 and September-October 2003. The GCDE consists of three rectangular pointing grids covering Galactic longitudes $|l| \le 30^\circ$ and Galactic latitudes $|b| \le 10^\circ$, and a fourth grid covering the same longitude range but $|b| \le 20^\circ$ in latitude. The GPS consists of pointings, following a saw-tooth pattern, along the Galactic plane within $|b| \le 6.4^\circ$. Details of the core program observing strategy can be found in \citet{winkler01}.

For each pointing and each of the 19 detectors, the SPI events data were gain corrected and binned into 0.25 keV wide bins \citep{knodlseder04}. The gain correction was performed for each orbit to account for long-term drifts that arise from temperatures variations and degradation of the detectors \citep{lonjou04}. 

\begin{figure}
	\includegraphics[height=7cm,width=8cm]{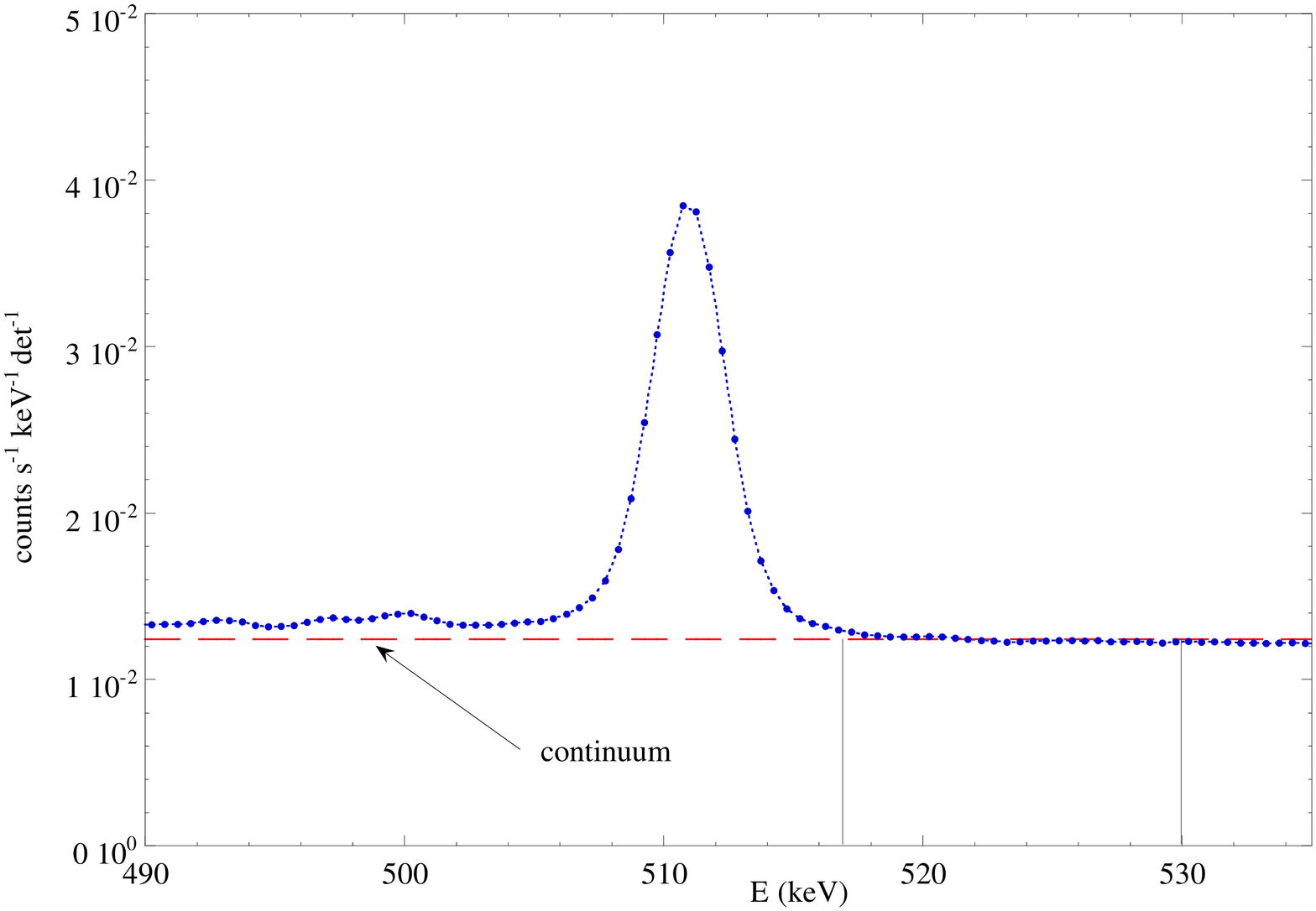}
	\caption{The instrumental background line at 511 keV. It is modelled with two components : one for the line and one for the continuum.}
	\label{fig:bck}
\end{figure}

Background modelling is a crucial point in this analysis because the instrumental line is roughly a hundred time stronger than the astrophysical one. We have modelled the background with two components (see Figure~\ref{fig:bck}): one for the line and one for the continuum \citep{jean03b, teegarden04, jean04}. 
The continuum component describes the temporal variation of the continuum level under the line which is considered to be flat. The "line" component contains the background line and the differences between the real continuum and the flat model.
We used the SPI background line identifications \citep{weidenspointner03}, the instrumental background simulations employing the MGGPOD Monte Carlo suite \citep{weidenspointner04a, weidenspointner04b} and others published informations on detectors to identify the main contributors to the 511~keV line and continuum around it. The contributions of the different isotopes (prompt and delayed) have been fitted to the data assuming that their production rate scales with the rate of the Germanium saturated events. Therefore, templates describing the temporal variations of the background components are generated.

\section{Model Fitting, background model and morphology}
\label{modelfitting}
\begin{figure}
	\includegraphics[height=6cm,width=8cm]{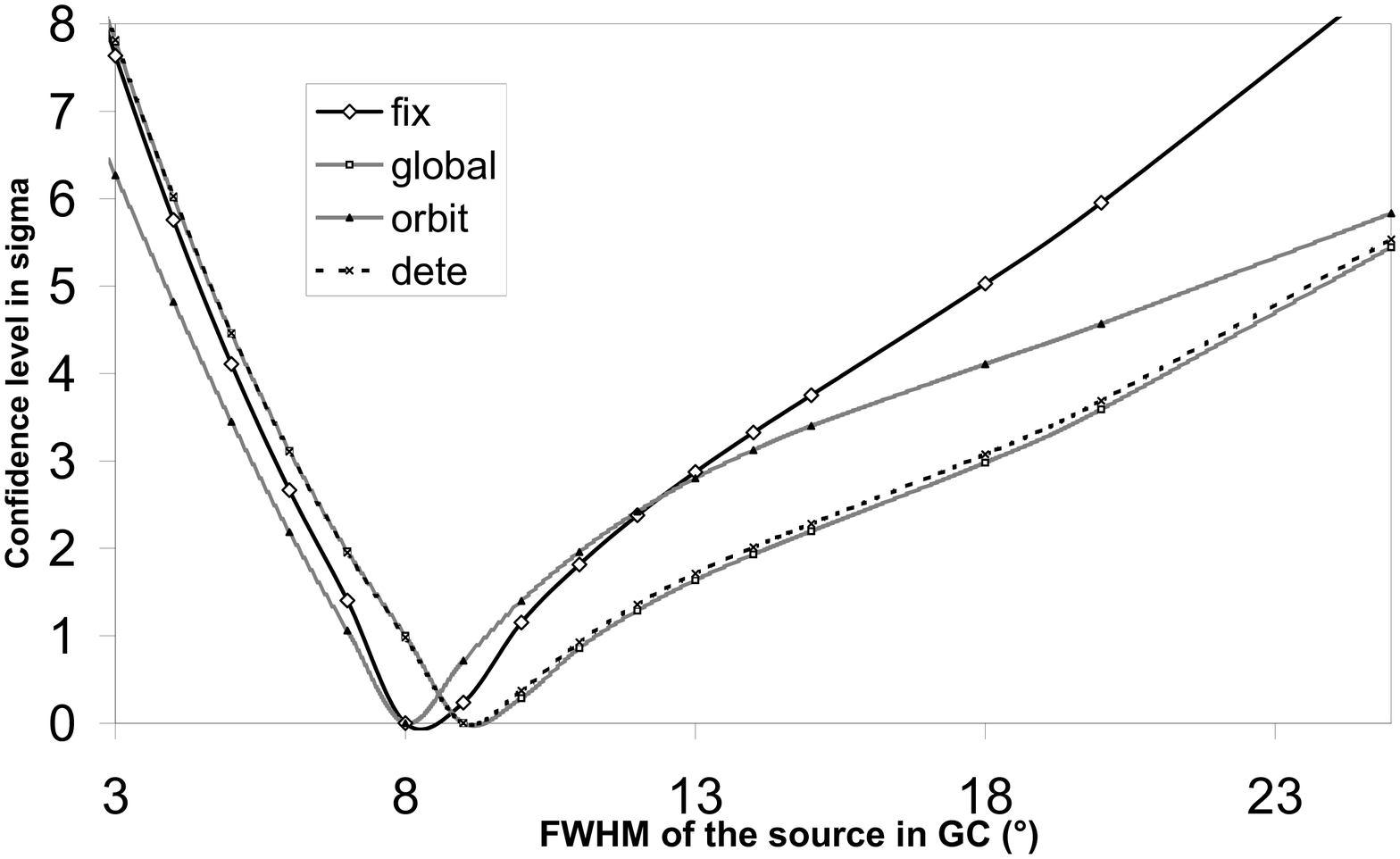}
	\caption{Confidence level as a function of the spatial distribution. The $\delta$ likelihood follows a $\chi2$ distribution, thus the confidence level is the square root of the $\delta$ likelihood because there is one degree of freedom.}
	\label{fig:model_fit}
\end{figure}

Using model fitting techniques \citep{knodlseder04}, we derived constraints on the morphology of the 511 keV line emission and to test our background models. We assume a 2D Gaussian centered on the Galactic Center for the spatial distribution, and we fit the flux in a 5 keV wide bin (508.5-513.5 keV). During the fit we can free some parameters in order to match the background model to the data. We have used four different fit types: fix (fixed background model), global (1 free parameter for the entire observation), orbit (1 free parameter per revolutio), dete (1 free parameter per detector).
The similarity of the results obtained with this four background fitting options (see Figure~\ref{fig:model_fit}) indicates the high quality of our background model. However, there remains room for improvement, thus this work continues.
A FWHM=$8^{+3 \circ}_{-2}$ is deduced with a 2$\sigma$ confidence level is found for the Gaussian. The effect of a disk component has been tested but was founded  to be smaller than the effect of the size of the bulge component. For more details on the morphological aspects see \citet{weidenspointner04c}.
\begin{figure}
	\includegraphics[height=6cm,width=8cm]{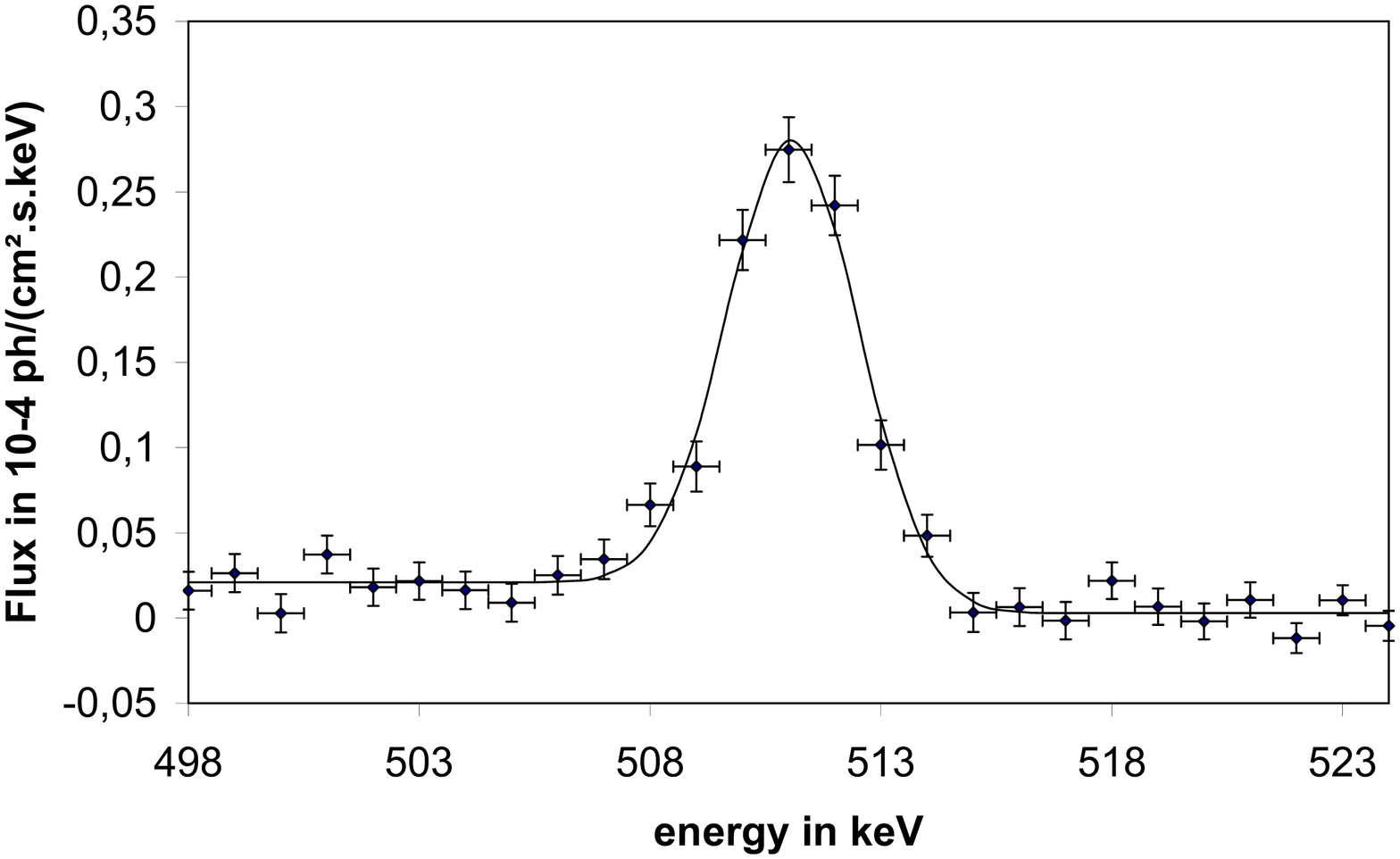}
	\caption{511 keV flux obtained using a model of the spatial distribution
	 consisting of a Gaussian centered on the Galactic Center with a FWHM of $8^{\circ}$.}
	\label{fig:spectre}
\end{figure}

\section{Results}
\label{results}
\begin{figure}
	\includegraphics[height=10cm,width=8cm]{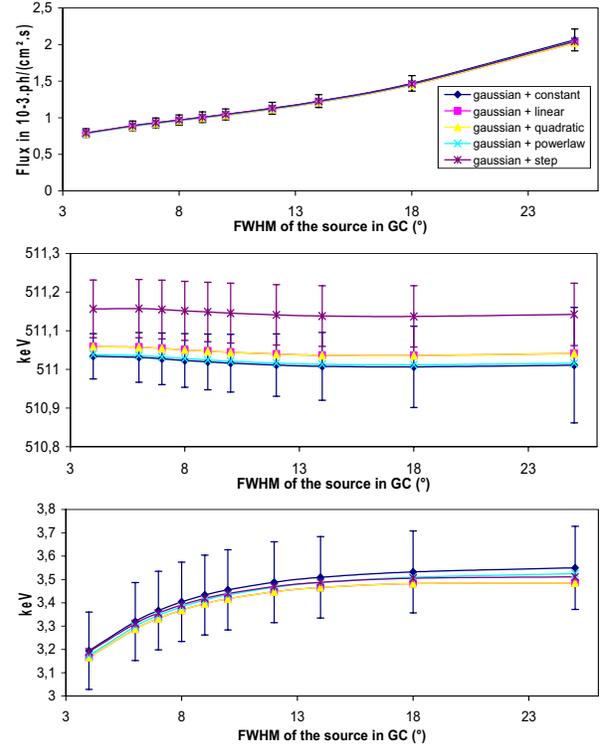}
	\caption{Flux, Position and Width of the 511 keV line from 
the Galactic Center as a function of the size of the spatial distribution.}
	\label{fig:line_features}
\end{figure}

We used model fitting techniques in order to extract the spectrum. In this case, we fit the same 2D Gaussian spatial model to the data in each 1 keV energy bin in the vicinity of the 511 keV line. We free one global parameter per energy bin in order to correct for detector degradation effects and to reduce systematic errors. As an example, the resulting photon spectrum obtained using a $8^{\circ}$ FWHM Gaussian for the spatial distribution (our best fitting model distribution) is shown in Figure~\ref{fig:spectre}. The spectrum has been fitted with a Gaussian to derive the flux, energy and the width of the line. The fit was performed with different models for the continuum under the line and for several widths of the spatial distributions. The results of this analysis are summarized in Figure~\ref{fig:line_features}.
Thus the uncertainties quoted for these results include the statistical uncertainties and the uncertainties on the spatial distribution. The width of the astrophysical line is the measured
width after deconvolving the instrumental width (FWHM$=2.10^{+0.02}_{-0.02}$~keV). We obtained a flux of $0.96^{+0.21}_{-0.14} \times 10^{-3}$ ph~cm$^{2}$~s$^{-1}$, a line centroid of $511.02^{+0.08}_{-0.09}$~keV and a width of the astrophysical line of $2.67^{+0.3}_{-0.33}$~keV. The flux value is strongly dependent on the size of the emission, but the position and width of the line are relatively stable. The best $\chi^2$ is obtained with a step function component in the background under the line.
Formally the continuum level to the left of the 511~keV line (see Figure ~\ref{fig:spectre}) is significantly higher than to its right. 
This is an indication of positronium continuum emission, but at this stage of the analysis we are not in a position to quote a definitive value for the positronium continuum flux and the positronium fraction.


\begin{figure}
	\includegraphics[height=10cm,width=8cm]{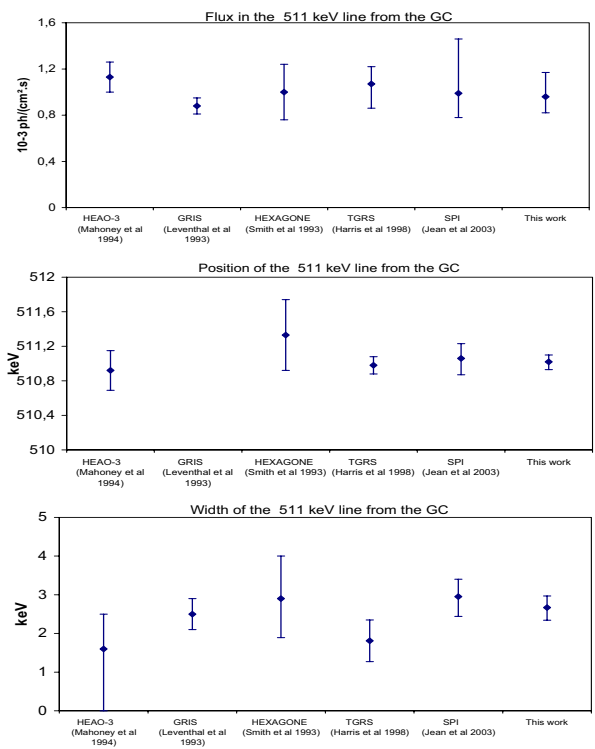}
	\caption{Flux, position and width of the GC 511 keV line measured by
	hight resolution spectrometers.}
	\label{fig:comparaison}
\end{figure}

\section{Conclusion}
\label{conclusion}
The analysis of SPI data that have been recorded during the first year INTEGRAL GCDE have provided constraints on the shape of the 511 keV line from the Galactic Center region. We derived a flux of  $0.96^{+0.21}_{-0.14} \times 10^{-3}$~ph~cm$^{-2}$~s$^{-1}$, a width of $2.67^{+0.3}_{-0.33}$~keV (FWHM) and a line centroid of $511.02^{+0.08}_{-0.09}$~keV. 
In general the results are consistent with earlier measurements (see Figure~\ref{fig:comparaison}), yet the measurement of the line width is formally inconsistent with the TGRS measurement on a 1 $\sigma$ confidence level. Such a significance for the energy and the width had never been obtained. One should note that the HEAO3 results are given for a point source at GC and that the GRIS results don't take into account the uncertainties on the spatial distribution, so a direct comparison is not possible.
Our analysis indicates the presence of the positronium continuum, but we cannot deduce any value for the positronium fraction because of systematic uncertainties.
These results can be used in order to determine and to understand the properties of the annihilation medium \citep{guessoum04}. We are now refining this work by improving the background modeling and including much data.

\section*{Acknowledgments}

Based on observations with INTEGRAL, an ESA project with instruments and science data centre funded by ESA member states (especially the PI countries: Denmark, France, Germany, Italy, Switzerland, Spain), Czech Republic and Poland, and with the participation of Russia and the USA.


\bibliographystyle{aa}
\bibliography{esapub}

\end{document}